\documentclass[11pt]{article}
\usepackage{amsfonts}

\newcommand{\set}[1]{\left\{ #1\right\}}
\newcommand{\gilt}{:}
\newcommand{\sodass}{\,:\,}
\newcommand{\setGilt}[2]{\left\{ #1\sodass #2\right\}}

\newcommand{\realrange}[2]{\left[#1, #2\right]}

\newcommand{\unitrange}[2]{\realrange{0}{1}}

\newcommand{\llabel}[1]{\label{\labelprefix:#1}}
\newcommand{\labelprefix}{} 

\newcommand{\discussionsize}{\small}

\newcommand{\frage}[1]{}

\newenvironment{code}{\noindent
\begin{tabbing}%
\hspace{2em}\=\hspace{2em}\=\hspace{2em}\=\hspace{2em}\=\hspace{2em}\=%
\hspace{2em}\=\hspace{2em}\=\hspace{2em}\=\hspace{2em}\=\hspace{2em}\=%
\kill}{\end{tabbing}}

\newcommand{\labelcommand}{}
\newcommand{\captiontext}{}
\newsavebox{\codeparam}
\newcounter{lineNumber}
\newenvironment{disscodepos}[3]{%
\renewcommand{\labelcommand}{#2}%
\renewcommand{\captiontext}{#3}%
\sbox{\codeparam}{\parbox{\textwidth}{#3}}%
\begin{figure}[#1]\begin{center}\begin{code}\setcounter{lineNumber}{1}}{%
\end{code}\end{center}\caption{\llabel{\labelcommand}\captiontext}\end{figure}}

{\end{disscodepos}}

\newcommand{\Is}       {:=}

\newdimen\endofsize\endofsize=0.5em
\def\endofbeweis{~\quad\hglue\hsize minus\hsize
                 \hbox{\vrule height \endofsize width
\endofsize}\par}

\usepackage[utf8]{inputenc}
\usepackage{algorithmic}
\usepackage{tabularx}
\usepackage{algorithm}
\usepackage{amsmath}
\usepackage{amssymb}
\usepackage{color}
\usepackage{fullpage}     
\usepackage{latexsym}
\usepackage{makeidx}
\usepackage{multicol}
\usepackage{numprint}
\usepackage{t1enc}
\usepackage{times}
\usepackage{graphicx}
\usepackage{url}
\usepackage[left=2cm,top=1.5cm,right=2cm]{geometry}
\npdecimalsign{.} 

\definecolor{mygrey}{gray}{0.75}
\newcommand{\ie}{i.e.\ }
\newcommand{\etal}{et~al.\ }
\newcommand{\eg}{e.g.\ }

\def\MdR{\ensuremath{\mathbb{R}}}

\newcommand{\mytitle}{ {\color{red}VieM} v1.00 -- {\color{red}Vie}nna {\color{red}M}apping and Sparse Quadratic Assignment \\ User Guide}
\begin{document}
\title{\mytitle}
\author{Christian Schulz and Jesper Larsson Träff\\ 
        \ \\
	\textit{University of Vienna},
	\textit{Vienna, Austria} \\
	\textit{Email: \url{christian.schulz}\url{@univie.ac.at}}  \\
        \ \\
	\textit{Technical University of Vienna}, \\
	\textit{Vienna, Austria} \\
	\textit{Email: \url{traff}\url{@par.tuwien.ac.at}} }
\date{}

\maketitle
\begin{abstract}
This paper severs as a user guide to the mapping framework VieM (Vienna Mapping and Sparse Quadratic Assignment). We give a rough overview of the techniques used within the framework and describe the user interface as well as the file formats used. 
\end{abstract}

\tableofcontents
\thispagestyle{empty}

\vfill
\pagebreak
\section{Introduction}
Communication performance between processes in high-performance
systems depends on many factors. For example, communication is
typically faster if communicating processes are located on the same
processor node compared to the cases where processes reside on
different nodes.  This becomes even more pronounced for large
supercomputer systems where processors are hierarchically organized
into, \eg islands, racks, nodes, processors, cores with corresponding
communication links of similar quality.  Given the communication
pattern between processes and a hardware topology description that
reflects the quality of the communication links, one hence seeks to
find a good mapping of processes onto processors such that pairs of
processes exchanging large amounts of information are located closely.

Such a mapping can be computed by solving a corresponding quadratic
assignment problem (QAP) which is a hard optimization problem.  Sahni
and Gonzalez~\cite{SahniG76} have shown QAP to be strongly NP-hard
and, unless P=NP, admitting no constant factor approximation
algorithm.  In addition, there are no algorithms that can solve
meaningful instances with $n>20$ to optimality in a reasonable amount
of time~\cite{burkard1998quadratic}.  Hence, heuristic algorithms are
necessary in order to solve large scale instances.  Multiple
heuristics have been proposed to tackle real world
instances~\cite{brandfass2013rank,heider1972computationally,muller2013optimale}.

We make two important assumptions that are typically
valid for modern supercomputers and the applications that run on
those.  First, communication patterns are almost always sparse since
not all processes have to communicate with each other.  This is
especially true for large scale scientific simulations in which the
underlying models of computation and communication are already sparse, see,
\eg~\cite{catalyuerek1996dis,heuvelinecoop,schloegel2000gph}.
To efficiently parallelize the simulation one normally employs graph
partitioning techniques which then in turn yield a sparse
communication pattern between the processes.  Second, we assume that
the hardware communication topology under consideration is
hierarchical with communication links on the same level in the
hierarchy featuring the same communication speed. This is typically
observed in current high-performance systems, \eg SuperMUC\footnote{Leibniz Supercomputing Centre, Gauss Centre for Supercomputing e.V.}.
Using these assumptions, we derive algorithms that are able to create high quality mappings, as well as faster local search algorithms for improving assignments. 

\textbf{Problem Definition}. The total communication requirement between the set of processes in an application can be modeled by a weighted communication graph. 
The underlying hardware topology can likewise be modeled by a weighted graph.
Our abstract problem
is to embed the communication graph onto the topology graph under optimization criteria that we explain below. We assume 
that the number of nodes in host and topology graphs are the same.
Unless otherwise mentioned, a processing element (PE) typically represents a core of a machine. 

Throughout the user guide, $\mathcal{C}\in \MdR^{n \times n}$ denotes the communication matrix and $\mathcal{D}\in \MdR^{n \times n}$ the topology matrix or distance matrix. More precisely, $\mathcal{C}_{i,j}$ describes the amount of communication that has to be done between process $i$ and $j$ and $\mathcal{D}_{i,j}$ represents the weighted distance between PE $i$ and PE $j$.
That is, the cost for communicating the amount $\mathcal{C}_{i,j}$ between processors $i$ and $j$ is $\mathcal{C}_{i,j}\mathcal{D}_{i,j}$.
We follow Brandfass~\etal~\cite{brandfass2013rank} and others, and model the embedding problem as a quadratic assignment problem (QAP): Find a one-to-one mapping $\Pi$ of processes to PEs which minimizes the overall communication cost. More precisely, we want to minimize 
$J(\mathcal{C},\mathcal{D}, \Pi) := \sum_{i,j} \mathcal{C}_{\Pi(i), \Pi(j)}\mathcal{D}_{i,j}$ 
where the sum is over all PE pairs and $k=\Pi(i)$ means that process $k$ is assigned to PE $i$. 
Our framework assumes that $\mathcal{C}$ and $\mathcal{D}$ are symmetric -- otherwise one can create equivalent QAP problems with symmetric inputs \cite{brandfass2013rank}.

Graph partitioning is a key component in our algorithms to find initial solutions. 
The \emph{graph partitioning problem} looks for \emph{blocks} of nodes $V_1$,\ldots,$V_k$ 
that partition $V$, \ie $V_1\cup\cdots\cup V_k=V$ and $V_i\cap V_j=\emptyset$
for $i\neq j$. The \emph{balancing constraint} demands that 
$\forall i\in 1..k\gilt c(V_i)\leq L_{\max}\Is (1+\epsilon)\lceil c(V)/k \rceil$ for
some parameter $\epsilon$. 
In the \emph{perfectly balanced case} the imbalance parameter $\epsilon $ is set to zero, \ie no deviation from the average is allowed.
One commonly used objective is to minimize the total \emph{cut} $\sum_{i<j}w(E_{ij})$ where 
$E_{ij}\Is\setGilt{\set{u,v}\in E}{u\in V_i,v\in V_j}$. A vertex $v \in V_i$ that has a neighbor $w \in V_j, i\neq j$, is a boundary vertex.

\vfill
\pagebreak

\section{Mapping Techniques within VieM}
We now give a rough overview over the algorithms implemented in our framework. For details on the algorithms, we refer the interested reader to the corresponding paper~\cite{DBLP:journals/corr/SchulzT17}. \\
\subsection{Local Search}
Heider~\cite{heider1972computationally} proposes a method to improve an already given permutation/mapping. 
The method repeatedly tries to perform swaps in the assignment. To do so, the author defines a pair-exchange neighborhood $N(\Pi)$ that contains all permutations that can be reached by swapping two elements in $\Pi$. Here, swapping two elements means that $\Pi(i)$ will be assigned to processor $j$ and $\Pi(j)$ will be assigned to processor $i$ after the swap is done. The algorithm then looks at the neighborhood in a cyclic manner. More precisely, in each step the current pair $(i,j)$ is updated to $(i,j+1)$ if $j<n$, to $(i+1,i+2)$ if $j=n$ and $i < n-1$, and lastly to $(1,2)$ if $j=n$ and $i = n-1$. A swap is performed if it yields  positive gain, \ie the swap reduces the objective. The overall runtime of the algorithm is $O(n^3)$. We denote the~search~space~with~$N^2$.
To reduce the runtime, Brandfass \etal~\cite{brandfass2013rank} introduce a couple~of~modifications. 
Initially computing as well as recomputing the objective function after a swap is performed is an expensive step in the algorithm. 
In their work, both the communication pattern as well as the distances between the PEs are given as complete matrices. 
These matrices have a quadratic number of elements and hence the initial computation of the objective function costs $O(n^2)$ time.
After a swap is performed, Brandfass~\etal update the objective using the objective function value before the swap. 
Overall, an update step in their algorithm takes $O(n)$ time which is clearly a bottleneck for sparse communication patterns. 
We described methods how we speed up the initial computation as well as the update of the objective. This yields much faster local search algorithms.

In addition to that we defined swapping neighborhoods using the communication graph $G_\mathcal{C}$. 
In the simplest version, assignments are only allowed to be swapped if the processes are connected by an edge in the communication graph, \ie the processes have to communicate with each other. 
We denote this neighborhood with $N_\mathcal{C}$. The size of the search space is $O(m)$ since it contains exactly $m$ pairs that may be swapped. 
Swaps are performed in random order. 
Local search terminates after $m$ unsuccessful swaps, \ie all pairs have been tried and no swap resulted in a gain in the objective.
Note that this approach assumes that swaps with positive gain are close in terms of graph theoretic distance in the communication graph. 
We also define augmented neighborhoods in which swaps are allowed if two processes have distance less than $d$ in the communication graph. We denote this neighborhood by $N^d_\mathcal{C}$. Note that this creates a sequence of neighborhoods increasing in size $N_\mathcal{C} \subseteq N^2_\mathcal{C} \subseteq \ldots \subseteq N^n_\mathcal{C} =  N^2$ where $N^2$ is the largest neighborhood used by Brandfass~\etal~\cite{brandfass2013rank}.

\subsection{Hierarchical Initial Solutions}
Our framework also contains algorithms to initially create solutions.
Throughout this section, we assume that the input communication matrix is already given as a graph $G_\mathcal{C}$, \ie no conversion of the matrix into a graph is necessary.
 More precisely, the graph representation is defined as $G_\mathcal{C}:=(\{1,\ldots, n\}, E[\mathcal{C}])$ where $E[\mathcal{C}] :=\{(u,v) \mid \mathcal{C}_{u,v} \not = 0\}$.
 In other words, $E[\mathcal{C}]$ is the edge set of the processes that need to communicate with each other. 
Note that the set contains forward and backward edges, and that the weights of the edges in the graph correspond to the entries in the matrix~$\mathcal{C}$. 

\label{s:mainsection}
\label{s:main}
Our most successful strategy is a top down approach. Intuitively, we want to identify subgraphs in the
communication graph of processes that have to communicate much with
each other and then place such processes closely, \ie on the same
node, same rack and so forth.  In the following, we assume a
homogeneous hierarchy of the supercomputer, but our algorithms can be extended to heterogeneous hierarchies in a straightforward way. Let $\mathcal{S}=a_1, a_2,
..., a_k$ be a sequence describing the hierarchy of the
supercomputer. The sequence should be interpreted as each processor having
$a_1$ cores, each node $a_2$ processors, each rack $a_3$ nodes, \ldots.

The \emph{top down approach} starts by computing a \emph{perfectly balanced} partition of $G_\mathcal{C}$ into $a_k$ blocks each having $n/a_k$ vertices (processes). The partitioning task is done using the techniques provided by Sanders and Schulz~\cite{kabapeE} which provide high quality partitions and guarantee that each block of the output partition has the specified amount of vertices. In principle, the nodes of each block will be assigned completely to one of the $a_k$ system entities. 
Each of the system entities provides precisely $n/a_k$ PEs. 
We then proceed recursively and partition each subgraph induced by a block into $a_{k-1}$ blocks and so forth. 
The recursion stops as soon as the subgraphs have only $a_1$ vertices left. 
In the base case, we assign processes to permutation~ranks.

\section{Graph Format}
\label{ss:graphformat}
\subsection{Input File Format}
The graph format used by our programs is the same as used by Metis \cite{karypis1998fast}, Chaco \cite{chaco} and the graph format that has been used during the 10th DIMACS Implementation Challenge on Graph Clustering and Partitioning~\cite{benchmarksfornetworksanalysis}. 
The input graph has to be undirected, without self-loops and without parallel edges.

To give a description of the graph format, we follow the description of the Metis 4.0 user guide very closely. A graph $G=(V,E)$ with $n$ vertices and $m$ edges is stored in a plain text file that contains $n+1$ lines (excluding comment lines). The first line contains information about the size and the type of the graph, while the remaining $n$ lines contain information for each vertex of $G$. Any line that starts with \% is a comment line and is skipped.

The first line in the file contains either two integers, $n$ $m$, or three integers, $n$ $m$ $f$. The first two integers are the number of vertices $n$ and the number of undirected edges of the graph, respectively. Note that in determining the number of edges $m$, an edge between any pair of vertices $v$ and $u$ is counted \emph{only once} and not twice, \ie we do not count the edge $(v,u)$ from $(u,v)$ separately. The third integer $f$ is used to specify whether or not the graph has weights associated with its vertices, its edges or both. If the graph is unweighted then this parameter can be omitted. It should be set to $1$ if the graph has edge weights, 10 if the graph has node weights and 11 if the graph has edge and node weights. However, note that since we compute one-to-one mappings, node weights are ignored.

The remaining $n$ lines of the file store information about the actual structure of the graph. In particular, the $i$th line (again excluding comment lines) contains information about the $i$th vertex. Depending on the value of $f$, the information stored in each line is somewhat different. In the most general form (when $f=11$, \ie we have node and edge weights) each line has the following structure:
\begin{center}
       $c\, v_1\, w_1\, v_2\, w_2 \ldots v_k\, w_k$ 
\end{center}
where $c$ is the vertex weight associated with this vertex, $v_1, \ldots, v_k$ are the vertices adjacent to this vertex, and $w_1, \ldots, w_k$ are the weights of the edges. Note that the vertices are numbered starting from 1 (not from 0). Furthermore, the vertex-weights must be integers greater or equal to 0, whereas the edge-weights must be strictly greater than 0.

\begin{figure}[h!]
\begin{center}
\includegraphics[width=.8\textwidth]{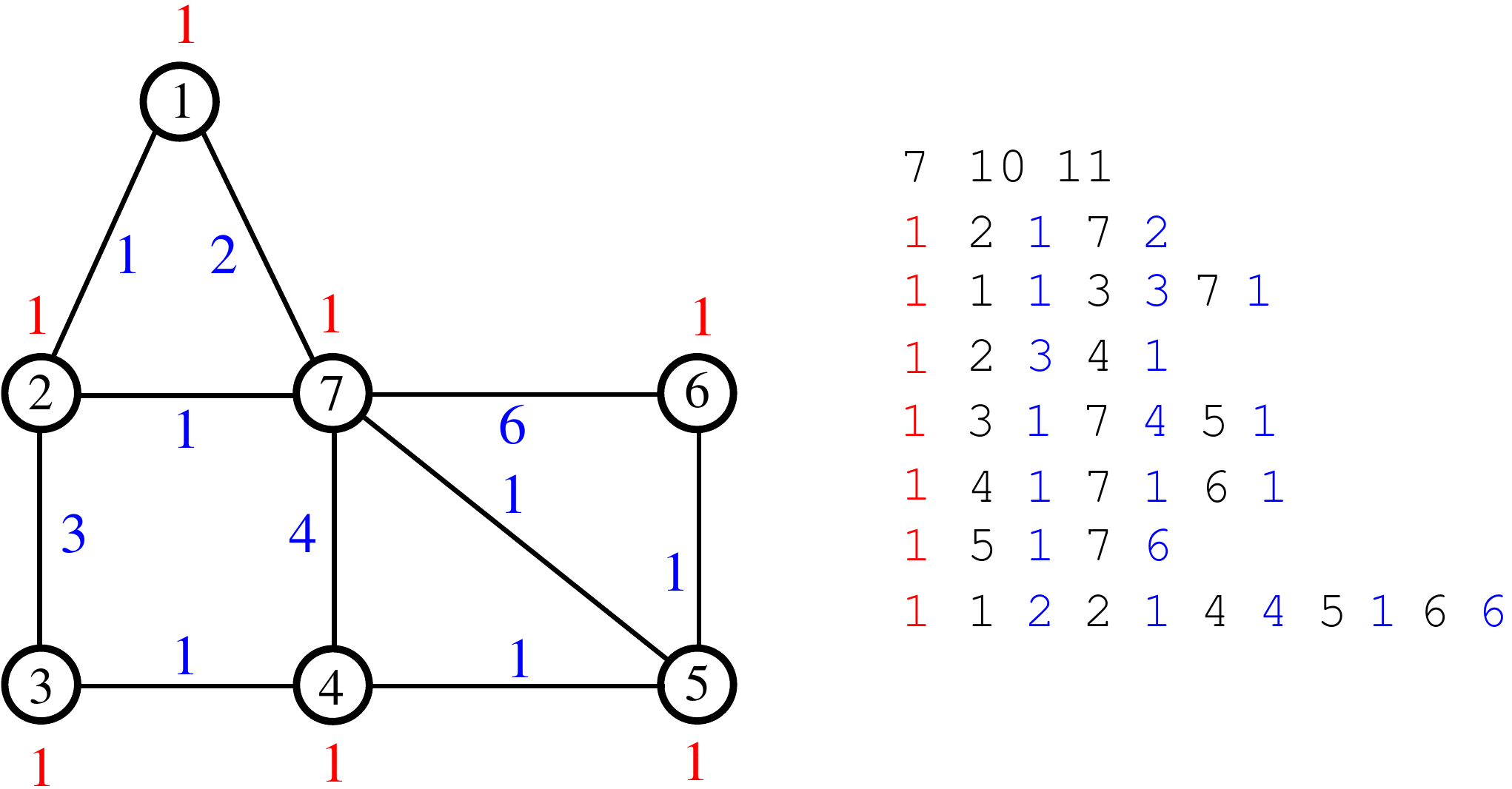}
\end{center}

\caption{An example graph and its representation in the graph format. The IDs of the vertices are drawn within the cycle, the vertex weight is shown next to the circle ({\color{red}red}) and the edge weight is plotted next to the edge ({\color{blue}blue}).}
\label{fig:example}
\end{figure}

\subsection{Output File Formats}
The output format of a mapping is basically a text file named \emph{permutation}. 
This file contains $n$ lines. 
In each line the mapped ID of the corresponding vertex is given, \ie line $i$ contains the mapped processor ID of the vertex $i$ (here the vertices are numbered from 0 to $n-1$).
The processor IDs are numbered consecutively from 0 to $n-1$. 
\subsection{Troubleshooting}
VieM should not crash! If VieM crashes it is mostly due to the following reasons: the provided graph contains self-loops or parallel edges, there exists a forward edge but the backward edge is missing or the forward and backward edges have different weights, or the number of vertices or edges specified does not match the number of vertices or edges provided in the file.
Please use the \emph{graphcheck} tool provided in our package to verify whether your graph has the right input format. If our graphcheck tool tells you that the graph that you provided has the correct format and VieM crashes anyway, please write us an email.

\vfill
\pagebreak
\section{User Interface}
Our package contains the following programs: viem, generate\_model, graphchecker, evaluator. To compile these programs you need to have Argtable, g++, and scons installed (we use argtable-2.10, g++-4.8.0, and scons-1.2). Once you have that you can execute \emph{compile.sh} in the main folder of the release. When the process is finished the binaries can be found in the folder \emph{deploy}. We now explain the parameters of each of the programs briefly.

\subsection{VieM}
\paragraph*{Description:} This is the mapping program. The default configuration of the program uses the top down approach as well as local search based on the communication graph with a neighborhood distances of 10. Note that the number of vertices in the model to be mapped has to be the same as the number of PEs specified using the hierarchy parameter string.
\paragraph*{Usage:\\} 
\begin{tabular}{ll}
viem &    [-{}-help] file [-{}-seed=<int>] [-{}-preconfiguration\_mapping]  -{}-hierarchy\_parameter\_string=<string>\\ 
     & -{}-distance\_parameter\_string=<string> [-{}-construction\_algorithm=<string>] \\
     &[-{}-distance\_construction\_algorithm=<string>] [-{}-local\_search\_neighborhood=<string>] \\
     & [-{}-communication\_neighborhood\_dist=<int>]
\end{tabular}
\subsection*{Options:\\}

\begin{tabularx}{\textwidth}{lX}
  file                                  & Path to file (model). \\
  -{}-help                      & Print help. \\ 
  -{}-seed=<int>                & Seed to use for the random number generator. \\
  -{}-preconfiguration\_mapping=<string> &    Use a preconfiguration for the partitioning tool within the mapping algorithm. One of strong, eco or fast. Default: eco. \\
  -{}-construction\_algorithm=<string>  &       Initial construction algorithm to use. One of random, identity, growing, hierarchybottomup, hierarchytopdown. Default: hierarchytopdown. \\
  -{}-distance\_construction\_algorithm=<string> & Construction algorithm to use to initially construct the distance matrix. Use one of hierarchy or hierarchyonline which does not store distance matrix. Default: hierarchy. \\
  -{}-hierarchy\_parameter\_string=<string> &   Specify hierarchy as 2:2:... for 2 cores per PE, 2 PEs per node, and so forth. \\
  -{}-distance\_parameter\_string=<string>  &   Specify distances between different levels as 1:10:... for 2 cores on the same PE have distance 1, and so forth \\
  -{}-local\_search\_neighborhood=<string>     & Local search neighborhood to use nsquare, nsquarepruned, or communication. Default: communication \\
  -{}-communication\_neighborhood\_dist=<int>  & set the communication neighborhood distance. Default: 10. \\
  -{}-output\_filename=<string> & Specify the output filename (default permutation). \\
\end{tabularx}
\vfill
\pagebreak
\subsection{Generate Model of Computation and Communication}
\paragraph*{Description:} This is program is for testing purposes. It takes a graph as input, partitions it using KaHIP~\cite{kaHIPHomePage,DBLP:journals/corr/SandersS13} and then creates a model of computation and communication. Here, blocks of the partition are vertices in the model and edge weights in the model are set to the number of edges that run between the respective blocks.
\paragraph*{Usage:\\} 

\begin{tabular}{ll}
generate\_model &   file -{}-k=<int> [-{}-help] [-{}-seed=<int>]  [-{}-preconfiguration=variant] \\
       &  [-{}-imbalance=<double>] [-{}-output\_filename=<string>] 
\end{tabular}
\subsection*{Options:\\}

\begin{tabularx}{\textwidth}{lX}
  file                        & Path to graph file that you want to partition and build the model from. \\
  -{}-k=<int>                   & Number of blocks to partition the graph into, i.e. number of vertices in the model. \\
  -{}-help                      & Print help. \\
  -{}-seed=<int>                & Seed to use for the random number generator. \\
  -{}-preconfiguration=variant & Use a preconfiguration for partitioning. (Default: eco) [strong| eco | fast | fastsocial| ecosocial| strongsocial ]. Strong should be used if quality is paramount, eco if you need a good tradeoff between partition quality and running time, and fast if partitioning speed is in your focus. Configurations with a social in their name should be used for social networks and web graphs. \\
  -{}-imbalance=<double>        & Desired balance. Default: 3 (\%). \\
  -{}-output\_filename=<string> & Specify the output filename (default model.graph). \\
\end{tabularx}
\vfill
\pagebreak

\subsection{Graph Format Checker}
\paragraph*{Description:} This program checks if the graph specified in a given file is valid. 
\paragraph*{Usage:\\} 
\begin{tabular}{ll}
graphchecker & file
\end{tabular}
\subsection*{Options:\\} 
\begin{tabularx}{\textwidth}{lX}
  file                       & Path to the graph file. \\
\end{tabularx}

\subsection{Evaluator}
\paragraph*{Description:} This program takes a model and a specification of the system hierarchy as well as a mapping of vertices of the model to processors in the system. It then computes the QAP objective. 
\paragraph*{Usage:\\} 
\begin{tabular}{ll}
evaluator & [-{}-help] file -{}-input\_mapping=<string> -{}-hierarchy\_parameter\_string=<string> \\
          & -{}-distance\_parameter\_string=<string>
\end{tabular}
\subsection*{Options:\\} 
\begin{tabularx}{\textwidth}{lX}
  -{}-help                              &      Print help. \\
  file                                  & Path to file (graph/model). \\
  -{}-input\_mapping=<string>            &     Input mapping to use. \\
  -{}-hierarchy\_parameter\_string=<string> &   Specify hierarchy as 2:2:... for 2 cores per PE, 2 PEs per node, and so forth. \\
  -{}-distance\_parameter\_string=<string>  &   Specify distances between different levels as 1:10:... for 2 cores on the same PE have distance 1, and so forth

\end{tabularx}

\vfill\pagebreak
\bibliographystyle{plain}
\bibliography{phdthesiscs}

\begin{thebibliography}{10}

\bibitem{benchmarksfornetworksanalysis}
D.~A. Bader, H.~Meyerhenke, P.~Sanders, C.~Schulz, A.~Kappes, and D.~Wagner.
\newblock Benchmarking for graph clustering and partitioning.
\newblock In {\em Encyclopedia of Social Network Analysis and Mining}, pages
  73--82. Springer, 2014.

\bibitem{brandfass2013rank}
B.~Brandfass, T.~Alrutz, and T.~Gerhold.
\newblock Rank reordering for {MPI} communication optimization.
\newblock {\em Computers \& Fluids}, 80:372--380, 2013.

\bibitem{burkard1998quadratic}
R.~E Burkard, E.~Cela, P.~M. Pardalos, and L.~S. Pitsoulis.
\newblock The quadratic assignment problem.
\newblock In {\em Handbook of combinatorial optimization}, pages 1713--1809.
  Springer, 1998.

\bibitem{catalyuerek1996dis}
{\"U}.~V. {\c{C}}ataly{\"u}rek and C.~Aykanat.
\newblock {Decomposing Irregularly Sparse Matrices for Parallel Matrix-Vector
  Multiplication}.
\newblock In {\em Proc. of the 3rd Intl. Workshop on Parallel Algorithms for
  Irregularly Structured Problems}, volume 1117, pages 75--86. Springer, 1996.

\bibitem{heuvelinecoop}
J.~Fietz, M.~Krause, C.~Schulz, P.~Sanders, and V.~Heuveline.
\newblock {Optimized Hybrid Parallel Lattice Boltzmann Fluid Flow Simulations
  on Complex Geometries}.
\newblock In {\em Proc. of Euro-Par 2012 Parallel Processing}, volume 7484 of
  {\em LNCS}, pages 818--829. Springer, 2012.

\bibitem{heider1972computationally}
C.~H. Heider.
\newblock A computationally simplified pair-exchange algorithm for the
  quadratic assignment problem.
\newblock Technical report, DTIC Document, 1972.

\bibitem{chaco}
B.~Hendrickson.
\newblock {Chaco: Software for Partitioning Graphs}.
\newblock {\url{http://www.cs.sandia.gov/~bahendr/chaco.html}}.

\bibitem{karypis1998fast}
G.~Karypis and V.~Kumar.
\newblock {A Fast and High Quality Multilevel Scheme for Partitioning Irregular
  Graphs}.
\newblock {\em SIAM Journal on Scientific Computing}, 20(1):359--392, 1998.

\bibitem{muller2013optimale}
H.~M{\"u}ller-Merbach.
\newblock {\em Optimale reihenfolgen}, volume~15 of {\em \"Okonometrie und
  Unternehmensforschung}.
\newblock Springer-Verlag, 1970.

\bibitem{SahniG76}
S.~Sahni and T.~F. Gonzalez.
\newblock P-complete approximation problems.
\newblock {\em J. {ACM}}, 23(3):555--565, 1976.

\bibitem{kaHIPHomePage}
P.~Sanders and C.~Schulz.
\newblock {KaHIP -- Karlsruhe High Qualtity Partitioning Homepage}.
\newblock {\url{http://algo2.iti.kit.edu/documents/kahip/index.html}}.

\bibitem{kabapeE}
P.~Sanders and C.~Schulz.
\newblock {Think Locally, Act Globally: Highly Balanced Graph Partitioning}.
\newblock In {\em 12th Intl. Sym. on Experimental Algorithms (SEA'13)}, LNCS.
  Springer, 2013.

\bibitem{DBLP:journals/corr/SandersS13}
Peter Sanders and Christian Schulz.
\newblock Kahip v0.53 - karlsruhe high quality partitioning - user guide.
\newblock {\em CoRR}, abs/1311.1714, 2013.

\bibitem{schloegel2000gph}
K.~Schloegel, G.~Karypis, and V.~Kumar.
\newblock {Graph Partitioning for High Performance Scientific Simulations}.
\newblock In {\em The Sourcebook of Parallel Computing}, pages 491--541, 2003.

\bibitem{DBLP:journals/corr/SchulzT17}
Christian Schulz and Jesper~Larsson Tr{\"{a}}ff.
\newblock Better process mapping and sparse quadratic assignment.
\newblock {\em CoRR}, abs/1702.04164, 2017.

\end{thebibliography}
\end{document}